# A study on Determinants of Dividend Policy and its Impact on Financial Performances: A Panel Data Analysis for Indian Listed Firms

Dr. N Suresh[1*], Pooja M[2*]

[1]*Professor, Faculty of Management and Commerce, Ramaiah University of Applied Sciences*
*Contact No: +919980995377*

[2]*MBA Student, Faculty of Management and Commerce, Ramaiah University of Applied Sciences*
*Contact No: +919008603483*

[1]nsuresh.ms.mc@msruas.ac.in, [2]pooja.manoharan25@gmail.com

***Abstract***: ***Determination of the correct mix of dividend and retained earnings and its effect on profitability has been a subject of controversy in financial management literature. This paper seeks to contribute to the on-going debate by examining the relationship between dividend pay-out policy and financial performance of 60 firms listed on the National Stock Exchange between 2009-2018. The Return on Assets (ROA) served as surrogate for the dependent variable, profitability, while Dividend Pay-out ratio proxied for dividend policy and was the only explanatory variable. Control variables include firm size, asset tangibility and leverage. Regression result reveals a positive and significant relationship between dividend pay-out policy (DPO) and firm performance (ROA). It is recommended that companies should endeavour to put in place robust dividend pay-out policy that would encourage investment in projects that give positive Net Present Value.***

## 1.Introduction

Every investment under taken by investors incorporates a sole purpose of maximizing wealth; and shareholders tend to speculate so as to create profit. Dividend is in the avenue through which investors in a company are rewarded for his or her investment. The primary objective of monetary management is that the maximization of the owners' wealth and to attain this, management as custodian of owners' interest, strives to make wealth and add value to the prevailing assets of the organization on behalf of its owners. To do this, they're faced with three important categories of deciding which are; financing, investment, and dividend decisions. These strategic decisions are crucial within the achievement of this primary objective and intrinsically, management must be tactful in their approach especially when it involves the payment of dividends since they have to not only consider the question of what proportion of the company's earnings are needed for investment, but also take into consideration the possible effect of their decisions on share prices. Dividend policy is an action plan put in by a firm to work out what proportion of its residual profits are paid to shareholders as dividends and the way much are held back as retained earnings and this has significant effect on the worth of shares since the worth of a share depends significantly on the number of dividend paid dead set shareholders. The Dividend yield ratio is a vital ratio in evaluating investment while the dividend pay-out ratio shows the proportion of the full residual profits that's distributed as dividend to shareholders.

**Theories of dividend policies:**

Dividend policy is an important area of corporate finance which must be analysed with a rigorous model. Many financial theorists developed different theoretical model to represent dividend policies. Major theories of dividend in financial management are as under:

**1. Walter's model:** The first theoretical model of dividend policy is Walter's theory on dividend policy which is the relevance concept of dividend. This notion of dividend policy articulate that a dividend decision of the company affects its valuation. The companies paying higher dividends have more value in comparison to the companies that pay lower dividends or do not pay at all.

**2. Gordon's model:** Gordon's theory on dividend policy is also significant. It rests upon the belief of the 'relevance of dividends' concept. It is also called as 'Bird-in-the-hand' theory that asserts that the current dividends are important in determining the value of the firm. Gordon's model is mathematical models to compute the market value of the company using its dividend policy. Gordon's model clearly relates the market value of the company to its dividend policy.

## 2. Review of Literature

1. (H. Kent Baker, 2019)- The purpose of this paper was to identify the dividend policy determinants of SriLankan firms and why they pay dividends. The evidence supports the pecking order, signalling, free cash flow, catering and outcome theories using both secondary and primary data and the bird-in-the-hand theory using survey data. The findings are useful not only for corporate decision makers in establishing an appropriate dividend policy but also for shareholders in making investment decisions.

2. (N. Jayantha Dewasiri, 2019)- The purpose of this paper was to identify the determinants of dividend policy in an emerging and developing market. Hence, past dividend decision or pay-out, profitability and investment opportunities are a common set of determinants with implications for both propensities to pay dividends and its pay-out. The findings support theories of dividends such as signalling, outcome, catering, life cycle, FCF and pecking order.

3. (Bassam Jaara, 2018)- This research investigates the determinant of dividend policy for a sample of non-financial companies in Jordan over the period 2005–2016. The impact of historical dividends always positive and significant and signposts that firm's trend of dividend pay-out rather than the random paying. Risk has a negative impact on the pay-out levels. The analysis was depending on some theories that affect the dividend policy such as: Dividends irrelevance theory, bird in hand theory, pecking order theory, agency problems and signalling theory.

4. (Werema, 2018)- An analysis of firm performance following dividend policy changes was conducted. the decision to decrease dividends reverses a declining trend of poor performance, and reduces financial leverage and liquidity problems. Finally, consistent with previous studies, the findings here are that the market reacts negatively to announcements of dividend decreases.

5. (S.M. Tariq Zafar, 2012)- The present study attempts to analyse the impact of dividend on shareholders wealth of eleven selected Indian banks listed and actively traded in National Stock Exchange (NSE). The first part of paper gives an insight about the dividend and its legal implications. The second part consists of data and their analysis which revealed the fact that there is significant impact of dividend policy on the shareholder's wealth in Indian banking companies.

6. (Marfo-Yiadom & Agyei, 2011)- The main thrust of this study is to find out the relationship between dividend policy and performance of banks in Ghana. The results reinforce earlier findings that leverage, size of a bank and bank growth enhance the performance of banks. The age factor presents mixed results. Generally, the result is in tandem with earlier studies that dividend policy has an effect on firm value.

7. (Mian Sajid Nazir, 2010)- Corporate dividend policy has been remained a heavily investigated issue in corporate finance. Fixed effect and random effect models have been applied on the panel data. The results found that dividend policy has a strong significant relationship with the stock price volatility in KSE.

8. (David J. Denis, 2008)- In the US, Canada, UK, Germany, France, and Japan, the propensity to pay dividends is higher among larger, more profitable firms, and those for which retained earnings comprise a large fraction of total equity. Outside of the US there is little evidence of a systematic positive relation between relative prices of dividend paying and non-paying firms and the propensity to pay dividends. Overall, these findings cast doubt on signalling, clientele, and catering explanations for dividends, but support agency cost-based lifecycle theories.

9. (SAMY BEN NACEUR, 2006)- The authors study the dividend policy of 48 firms listed on the Tunisian Stock Exchange during the period 1996–2002. However, neither the ownership concentration nor the financial leverage seems to have any impact on dividend policy in Tunisia. Also, the liquidity of stock market and size negatively impacts the dividend payment. The results are somewhat robust to different specifications.

10. (Myers, 2004)- The importance of dividend cash flow as a signalling device to stockholders is also evident in the sample since even with high growth, the firm is willing to increase debt to fund increasing dividends. The firms in the sample desire to "put their money where their mouth is" by sending a strong positive signal to institutional owners to enhance reputation and maintain access to capital.

Dividend policy is most focused research area in finance. Although a lot of work has been done throughout the world about dividend policy, but still it is puzzle in finance. Globally, empirical literature showed diversified findings regarding relationship between financial performance and dividend policy.
The main objective of this study-
a) To examine the relationship between dividend pay-out and firm's financial performance
b) To appraise the effects of dividend payments on the changes in shareholders' funds

## 3. Data Sources and Methodology

### 3.1. Data sources

The determinants of dividend policy of companies observed the quantitative empirical research design. This form of research design has become critical due to the fact the researcher accrued numeric facts of variables from the financial statements for 10 consecutive years of the sampled companies for the study. This research will attempt to collect the financial data of 60 listed companies in India for the period covering the years 2009–2018.

### 3.2. Methodology

In the first objective, the selected Dependent variable is Return on Assets and the independent variables are Log of Total Assets, Asset Tangibility, Leverage, Dividend per share. Based on these variables Correlation and Panel OLS Regression analysis has been done.

In the second objective, the selected Dependent variable is Market capitalization and the independent variables are Total Assets and Dividend. Based on these variables Panel OLS Regression analysis has been done.

## 4. Results and Analysis

**Objective 1: Relationship between dividend pay-out and firm's financial performance**

**Table 1: Panel OLS Regression results.**

| Mode: Panel OLS Method | | | R Square: 0.785668 | |
|---|---|---|---|---|
| Dependent Variable: ROA | | | | |
| Method: Least square | | | Adjusted R Square: 0.760476 | |
| Independent Variable | Co-efficient | Standard Error | T-Statistic | Prob. |
| Constant | 0.283551 | 0.036276 | 70816404 | 0.0000 |
| AT | -0.028656 | 0.019027 | -1.506106 | 0.1326 |
| DIVSH | 0.001254 | 0.000245 | 5.124678 | 0.0000 |
| LEV | -0.147550 | 0.018777 | -7.858150 | 0.0000 |
| LOG OF TA | -0.034248 | 0.008757 | -3.911100 | 0.0001 |

**Source:** Based on E views output, Author analysis

Above Table 1. showing the results of Panel OLS regression for selected variables. Panel OLS Regression equation had been run between the 4 selected variables namely AT, DIVSH, LEV, Log of TA. Variables taken for this test is statistically significant in this case.

From the above table of Panel OLS regression results that the first variable
AT (Asset Tangibility) and the Probability of 0.1326 and T Statistics of -1.506106 this indicates that any changes in AT will not impact on the ROA(Profitability).
In the case of DIVSH (Dividend per share), the test shows us that there is a Probability of 0.0000 and T Statistics of 5.124678. It means that any changes in the DIVSH (Dividend per share) will impact positively on ROA(Profitability) of the firm.

In the case of LEV (Leverage), the probability of 0.0000 and T-Statistics of -7.858150. it means that any changes in LEV (Leverage) will affect the ROA(Profitability) negatively. Increase in LEV (Leverage) will decrease ROA(Profitability) and decrease in LEV (Leverage) will increase in ROA(Profitability).

In the case of Log of TA (Firm size), the probability of 0.0001 and T-Statistics of -3.911100. it means that any changes in Log of TA (Firm size) will affect the ROA(Profitability) negatively. Increase in Log of TA (Firm size) will decrease ROA(Profitability) and decrease in Log of TA (Firm size) will increase in ROA(Profitability).

**Objective 2: Appraise the effects of dividend payments on the changes in shareholders fund**

**Table 2: Panel OLS Regression results.**

| Mode: Panel OLS Method | | | R Square: 0.924865 | |
|---|---|---|---|---|
| Dependent Variable: MCAP | | | | |
| Method: Least square | | | Adjusted R Square: 0.916346 | |
| Independent Variable | Co-efficient | Standard Error | T-Statistic | Prob. |
| Constant | 10985.94 | 1723.412 | 6.374527 | 0.0000 |
| DIV | 29.17820 | 1.432493 | 20.36883 | 0.0000 |
| TA | 0.320084 | 0.031152 | 10.27493 | 0.0000 |

**Source:** Based on E views output, Author analysis

Above Table 2, showing the results of Panel OLS regression for selected variables. Panel OLS Regression equation had been run between the 2 selected variables namely DIV (Actual dividends paid), TA (Total Assets). Variables taken for this test is statistically significant in this case.
While testing this test there are some variables running out of the regression and such variables are not considered for this study.

From the above table of Panel OLS regression results, the variable DIV (Actual dividends paid) has a Probability of 0.0000 and T-Statistics of 20.36883. This indicates that any changes in DIV (Actual dividends paid) will impact on MCAP (Shareholders wealth) positively. It means that any changes in DIV (Actual dividends paid) will affect MCAP (Shareholders wealth).

In case of the variable TA (Total Assets) has a Probability of 0.0000 and T-Statistics of 10.27493. This indicates that any changes in TA (Total Assets) will impact on MCAP (Shareholders wealth) positively. It means that any changes in TA (Total Assets) will affect MCAP (Shareholders wealth).

**HYPOTHESIS TEST RESULTS**

**Table 3. HYPOTHESIS TEST RESULTS for Objective 1**

| Sl. No | Hypothesis Variable | $H_a$ | $H_0$ | P Value | Null Hypothesis Accept/Reject |
|---|---|---|---|---|---|
| 1 | Asset Tangibility | $\beta_1 = 0$ | $\beta_1 \neq 0$ | 0.1326 | Accept |
| 2 | Dividend Per Share | $\beta_2 = 0$ | $\beta_2 \neq 0$ | 0.0000** | Reject |
| 3 | Leverage | $\beta_3 = 0$ | $\beta_3 \neq 0$ | 0.0000** | Reject |
| 4 | Log of Total Assets | $\beta_4 = 0$ | $\beta_4 \neq 0$ | 0.0001* | Reject |

- The P value for the selected independent variable Asset Tangibility is 0.1326 which is more than 0.05. This implies that Asset Tangibility does not impact the Return on Assets. Hence, the Asset Tangibility for the selected period is not statistically significant. So, the Null Hypothesis $H_0$ is Accepted.

- The P value for the selected independent variable is Dividend Per Share is 0.0000 which is less than 0.05. This implies that Dividend Per Share impacts the Return on Assets. Hence, the Dividend Per Share for the selected period is statistically significant. So, the Null Hypothesis $H_0$ is Rejected.

- The P value for the selected independent variable is Leverage is 0.0000 which is less than 0.05. This implies that Leverage impacts the Return on Assets. Hence, the Leverage for the selected period is statistically significant. So, the Null Hypothesis $H_0$ is Rejected.

- The P value for the selected independent variable Log of Total Assets is 0.0001 which is less than 0.05. This implies that Log of Total Assets impacts the Return on Assets. Hence, the Log of Total Assets for the selected period is statistically significant. So, the Null Hypothesis $H_0$ is Rejected.

**Table 4. HYPOTHESIS TEST RESULTS for Objective 2**

| Sl. No | Hypothesis Variable | $H_a$ | $H_0$ | P Value | Null Hypothesis Accept/Reject |
|---|---|---|---|---|---|
| 1 | Dividends | $\beta_1 = 0$ | $\beta_1 \neq 0$ | 0.0000** | Reject |
| 2 | Total Assets | $\beta_2 = 0$ | $\beta_2 \neq 0$ | 0.0000** | Reject |

- The P value for the selected independent variable is Dividends is 0.0000 which is less than 0.05. This implies that Dividends impacts the Market Capitalization. Hence, the Dividends for the selected period is statistically significant. So, the Null Hypothesis $H_0$ is Rejected.

- The P value for the selected independent variable Total Assets is 0.0000 which is less than 0.05. This implies that Total Assets impacts the Market Capitalization. Hence, the Total Assets for the selected period is statistically significant. So, the Null Hypothesis $H_0$ is Rejected.

## 5. Conclusion and Future Recommendations

Dividends are very important for firms and country's economy. This paper examined the determinants of dividend policy on Indian firms for the period of 10 years from 2009-2019. Dividends are prone to more and more fluctuations like macro and micro economic factors that could impact dividends structure of firms. In the second objective, to study relation between dividend pay-out and firms performance the Panel OLS regression was conducted and it was found that out of 4 variables Asset Turnover Ratio, Dividend per share, Leverage and Log of Total assets are affected as these variables are statistically significant and determine the dividend policies of firms and there was a bi-directional relationship between Return on Asset (profitability) and Asset Tangibility, between Leverage and Return on Asset, between Log of Total assets and Return on Assets. In third objective, to find out effects of dividend payments on changes in shareholders fund the Panel OLS regression was performed and it was found that out of selected variables only 2 variables were affecting shareholders fund by payments of dividend.

Based on the aforementioned results, the study recommends that organizations should ensure that they have a good and robust dividend policy in place. This will enhance their profitability and attract investments to the organizations. Management of companies should also invest in projects that give positive Net Present Values, thereby generating huge earnings, which can be partly used to pay dividends to their equity shareholders. This will encourage investors (both local and foreign) to have stake in more firms that pay dividends consistently.

Limitations of the study were only 10years data was considered for the study. The study only had Indian companies and did not include the Foreign companies like MNC's. The study is limited to the country India.

Future work suggestions can be to consider more companies and make the study in large-scale, consider a large period of time study and also to include more variables for the study.